\documentclass{article}

\usepackage[accepted]{icml2026}

\usepackage{microtype}
\usepackage{graphicx}
\usepackage{subfigure}
\usepackage{booktabs}
\usepackage{hyperref}
\usepackage{xcolor}
\usepackage{amsmath}
\usepackage{amssymb}

\icmltitlerunning{AI Agents for Trustworthy Science}

\begin{document}

\twocolumn[
\icmltitle{Position: The Age of AI Agents Demands A New Scientific Paradigm To Sustain Trustworthy Science}

\icmlsetsymbol{equal}{*}

\begin{icmlauthorlist}
\icmlauthor{Belinda Mo}{lhr}
\end{icmlauthorlist}

\icmlaffiliation{lhr}{Long Horizon Research}
\icmlcorrespondingauthor{Belinda Mo}{belinda@longhorizonresearch.com}

\vskip 0.3in
]

\printAffiliationsAndNotice{}

\begin{abstract}
AI systems are becoming autonomous research agents that generate hypotheses, design experiments, and produce discoveries at scales beyond human oversight. As seen by increased submissions to ML venues, the verification gap between scientific output and our ability to check it is already widening, and autonomous agents make it worse by magnitudes given human-agent asymmetry. We argue that science must evolve its verification infrastructure, as it has before with peer review. However, while historical adaptations assumed human contributors who could be questioned and sanctioned, AI agents break this assumption. We propose criteria for an adapted verification infrastructure that emphasizes observable-by-default workflows, scalable verification, and clear attribution. We argue that without adaptation, ML and any scientific domain using agents face dangerous failures: experimental results that no person can verify, optimization for metrics over understanding, and accountability vacuums that erode scientific trust.
\end{abstract}

\section{Introduction}

AI systems are no longer just tools that scientists use. They are increasingly autonomous agents that generate hypotheses, design experiments, and produce discoveries at scales beyond human oversight. AlphaFold has predicted 200 million protein structures and won the 2024 Nobel Prize in Chemistry \citep{jumper2021alphafold}. AlphaEvolve discovered the first improvement to matrix multiplication in 56 years \citep{deepmind2025alphaevolve}. Sakana's AI Scientist generated complete research papers and passed peer review for a workshop in ICLR 2025 \citep{lu2024aiscientist}. 

These systems share a defining feature: they reason autonomously, making thousands of decisions without human involvement. Our existing methods for verifying experiments do not account for such a system. 

\textbf{We argue that the scientific method must evolve to address three interconnected challenges that AI agents create: observability (can we see what happened?), attribution (who is responsible?), and reproducibility (can we verify the results?).} These three mechanisms  break down when research contributions happen alongside agents that scale rapidly and cannot be held accountable.

\subsection{The Verification Gap Is Already Wide, and Agents Will Make It Worse}

The replication crisis revealed that scientific verification was already failing under human-speed science. Reproducibility is a defining feature of science, yet in 2015 it was found that only 36\% of psychology studies could be replicated with statistically significant results \citep{osc2015}. Nature found that across 1500 scientists, over 70\% of researchers had failed to reproduce others' experiments \citep{baker2016reproducibility}. In machine learning, benchmark results often fail to replicate when hyperparameters are properly controlled \citep{pineau2021reproducibility}.

The verification gap is the distance between scientific output and our ability to verify it, and agents will make this gap dramatically worse.

METR's research shows that the length of tasks AI agents can complete with 50\% reliability has doubled every seven months, accelerating to every four months in 2024-2025 \citep{metr2025longtasks}. \textbf{If agentic capabilities continue to doubling every 4 to 7 months while human verification capacity remains constant, simple extrapolation suggests the gap could grow by a factor of $2^{8}$ to $2^{15}$ (roughly 250--30,000$\times$) within five years.} Even with significant slowdown in capability growth, the asymmetry is stark.

\subsection{Three Challenges for Human-AI Science}

Agent-enabled discovery creates three interconnected challenges. These extend the reproducibility crisis framework \citep{munafo2017manifesto} to account for contributors who cannot be meaningfully questioned or held accountable.

\textbf{Observability: Can we see what happened?} In control theory, a system is ``observable'' if its internal state can be inferred from outputs \citep{kalman1960}. By this definition, AI agent reasoning is unobservable: we cannot reliably infer internal states. Agent observability can be improved via logs, metrics, and traces \citep{sridharan2018observability}. However, current agent-assisted research lacks all three. We do not face this issue when working with other human researchers, because we can substitute direct observation with \textit{social observation}: lab meetings, mentorship, questioning. With agents, not only are the internal states unreliable to infer, agents can also run for days or weeks autonomously, which can take too long for a scientist to review. 

\textbf{Attribution: Who is responsible?} Agents cannot be held accountable. They have no careers, reputations, or incentives. For human science, \textit{social accountability} (reputation, sanctions, career consequences) constrains even imperfect explanations. You can question a human collaborator and expect their answers to be constrained by consequences for being wrong. AI agents can be questioned, but their responses may not reflect actual reasoning processes \citep{huang2023hallucination,bordt2025xaistatistics}, and there are no consequences for unreliable answers. We need technical attribution mechanisms.

\textbf{Reproducibility: Can we verify results?} The reproducibility crisis predates AI, but agents introduce new failure modes: model drift, stochastic outputs, prompt sensitivity, and undocumented configurations. Current documentation practices capture none of this.

These three challenges share a common cause: \textbf{human science relies on social mechanisms for verification; AI science breaks those mechanisms}. When an agent explores 10,000 configurations overnight, documentation cannot be retrospective. When agents run autonomously for weeks, social observation fails. When contributors cannot be sanctioned, social accountability is meaningless.

\subsection{Science Has Adapted Before, But Previous Adaptations Assumed Human Contributors}

Science has evolved its verification infrastructure before. Peer review itself is 
remarkably recent: \textit{Nature} made external refereeing mandatory only in 
1973 \citep{baldwin2015}. Big Science developed contribution statements when papers 
began listing thousands of authors \citep{aad2015higgs,brand2015credit}.

However, peer review is already straining. Major ML venues receive tens of thousands of submissions; qualified reviewers cannot scale with volume. The social trust mechanism that peer review relies on, namely that reviewers and authors can question each other and that reputations constrain behavior, breaks down at scale even with human contributors. \textbf{When contributors are agents that run autonomously for days or weeks, social trust mechanisms fail entirely.}

Each adaptation solved a specific problem while preserving a common assumption: 
human contributors who can be questioned and held accountable.

\begin{itemize}
\item \textbf{Statistical methods} replaced intuitive judgment with formal inference, 
but the reasoning remains inspectable; a trained statistician can verify whether 
conclusions follow from data.
\item \textbf{Big Science} strained attribution, but every contributor can still 
explain their role and face consequences for misconduct.
\item \textbf{Peer review} enabled strangers to evaluate work, but assumes authors 
can clarify their reasoning when questioned.
\end{itemize}

These adaptations substituted \textit{technical mechanisms} for \textit{personal 
knowledge}: you don't need to know an author personally to trust peer-reviewed 
work. But \textit{social accountability} remained: authors can be questioned, 
sanctioned, and excluded.

\textbf{AI agents remove this backstop.} When an agent runs autonomously for days 
or weeks, there is no contributor to question. Explanations may not reflect actual 
reasoning \citep{huang2023hallucination}. There are no reputational consequences 
for unreliable outputs. The social mechanisms that previous adaptations preserved 
no longer apply.

Previous adaptations also took decades. We may not have decades; AI capabilities 
advance monthly.

\section{The Human-Agent Asymmetry}

Human-AI collaboration differs from human-human collaboration in ways that matter for verification:

\begin{table*}[t]
\centering
\small
\begin{tabular}{lll}
\toprule
\textbf{Dimension} & \textbf{Humans} & \textbf{AI Agents} \\
\midrule
Sustained attention & Hours; fatigue-limited & Days/weeks; compute-limited \\
Working memory & $\sim$4 items & Large context windows \\
Long-term memory & Lifespan & Session-bounded (currently) \\
Parallel exploration & Minimal; sequential & Massive; thousands of branches \\
Goal-setting & Intrinsic; value-driven & Delegated; prompt-dependent \\
Accountability & Can be held responsible & Cannot in meaningful sense \\
Explicability & Can explain (imperfectly) & Cannot verify explanations \\
\bottomrule
\end{tabular}
\caption{Key asymmetries between human and AI research collaborators.}
\label{tab:asymmetry}
\end{table*}

These differences invert the traditional research cycle. A human researcher spends weeks developing intuition about a problem, days running experiments, hours interpreting results. An agent inverts this: minutes generating hundreds of hypotheses, hours running thousands of experiments, requiring human judgment only to assess what matters.

This inversion creates a fundamental problem: \textbf{the work that most needs checking is the work that is hardest to check}.

Recent agents reason across domains \citep{wei2022chainofthought,bubeck2023sparks} and take autonomous actions \citep{yao2023react,wang2024llmagents}. DeepMind's knot theory collaboration illustrates the paradigm: neural networks identified patterns suggesting conjectures; human mathematicians proved theorems \citep{davies2021}. The network's reasoning was opaque, but mathematical proof established results. In fields without such proof mechanisms, we need alternative verification infrastructure.

AI agents also introduce reproducibility challenges that current documentation practices do not address: model drift (APIs change silently), stochastic outputs (different results each run), prompt sensitivity, and undocumented configurations. Methods sections have no conventions for documenting AI interactions.

\subsection{AI and the Verification Gap}

Luo et al.\ \citep{luo2025aiscientistpitfalls} systematically evaluated two prominent open-source AI Scientist systems and identified four failure modes: inappropriate benchmark selection (cherry-picking favorable datasets), data leakage (training-evaluation overlap), metric misuse, and post-hoc selection bias (similar to p-hacking). Their key finding: these failures ``can be easily overlooked in practice'' when examining the final paper alone. \textbf{Access to trace logs and code from the full automated workflow was necessary to detect the problems.}

AlphaFold predictions, despite Nobel Prize recognition, show significant discrepancies with experimental validation. A 2025 analysis of 74 GPCR structures found that ``while AlphaFold3 accurately captured global receptor architecture...its ligand positioning was highly variable and often inaccurate, rendering predictions unreliable'' \citep{alphafold2025gpcr}. Even high-confidence predictions contain errors not evident from confidence scores \citep{terwilliger2024alphafold}. Experimental validation remains essential, but as prediction volume vastly exceeds experimental capacity, which predictions get validated?

LLM-generated code introduces subtle bugs that propagate through research pipelines. A 2025 empirical study found that 35\% of LLM-generated code is less robust than human-written code, with over 90\% of deficiencies caused by missing conditional checks \citep{liu2025llmbugs}. In research contexts, such bugs may not cause obvious failures but instead produce silently incorrect results.

\section{Criteria for Adapted Verification}

Any adequate response to human-agent collaboration must satisfy certain criteria. The new scientific paradigm must preserve verification's epistemic function (ensuring claims are likely true), social function (maintaining institutional trust), and practical function (enabling others to build on work). However, they may be satisfied via different mechanisms.

\subsection{Discovery vs. Justification}

Hans Reichenbach \citep{reichenbach1938} distinguished the \textit{context of discovery} from the \textit{context of justification}. Discovery is how scientists actually arrive at ideas, through intuition, accidents, dreams, or opaque processes. Justification is how claims are evaluated and warranted, through evidence, logic, and reproducibility.

The history of science is full of discoveries from strange places. Kekul\'{e} claimed to discover the structure of benzene in a dream. Penicillin was discovered through via an accident. Many breakthroughs came from intuitions their authors could not articulate. What made these discoveries \textit{scientific} was not their origin but rather the rigor of their subsequent justification.

This matters for AI. \textbf{We must ensure that discoveries can be \textit{justified}}, traced through evidence chains that humans can evaluate, regardless of how they arose. The internal activations of an LLM that produces a discovery need not be interpretable for this purpose.

\textbf{Why do we accept opaque human reasoning but not opaque AI reasoning?} One might object: humans cannot fully explain their reasoning either. We cannot extract the full thought process from biological neural activity any more than from artificial neural networks. Yet we accept human explanations. Why?

The answer is social mechanisms. We accept human explanations because:
\begin{itemize}
\item \textbf{Reputation constrains confabulation:} Humans who provide unreliable explanations lose credibility over time.
\item \textbf{Accountability enables sanctions:} Humans who commit misconduct can be excluded from the scientific community.
\item \textbf{Questioning enables probing:} We can ask follow-up questions, challenge assumptions, request clarifications.
\end{itemize}

AI agents lack all three. They have no reputation to protect, cannot be sanctioned, and their responses to questioning may not reflect actual reasoning \citep{huang2023hallucination}. \textbf{Observable workflows are the technical substitute for social accountability}, as they constrain AI contributions the way reputation and sanctions constrain human contributions.

DeepMind's knot theory collaboration illustrates this \citep{davies2021}. Neural networks identified patterns suggesting conjectures, and mathematical proof justified them. The network's opacity was irrelevant because mathematical proof provided the technical mechanism for justification. In fields without such proof mechanisms, observable workflows must play an analogous role.

Peer review itself is scaled justification: strangers evaluating whether written justifications are adequate. The problem with human-AI science is that peer reviewers cannot adequately evaluate AI-generated justifications because: (1) the justifications may not reflect actual AI reasoning, (2) the scale of AI output exceeds human review capacity, and (3) AI-specific failure modes cannot be detected with currently available observability. 

\subsection{Criteria Any Solution Must Meet}

Any adequate scientific method for human-agent collaboration must satisfy several criteria. These do not specify implementation; they specify what any implementation must achieve.

\textbf{Observable-by-default workflows}: Documentation must happen automatically as work proceeds, not retrospectively. The ratio of documentation effort to experimental effort must scale with AI speed, requiring automated capture.

\textbf{Justification-preserving records}: Records must preserve evidence chains (which data supported which conclusions, which experiments tested which hypotheses) even when discovery processes are opaque.

\textbf{Scalable verification}: Human review cannot inspect every agent decision. We need tiered verification that focuses human attention where it matters most.

\textbf{Traceable attribution}: Attribution must identify who (or what) contributed specific elements while remaining tractable for accountability.

\textbf{Reproducibility infrastructure}: Documentation must capture AI-specific variability: model versions, prompts, configurations, random seeds.

\textbf{Failure-mode awareness}: Methods must address known AI failure modes: reward hacking, deceptive alignment, model drift, prompt sensitivity.

These criteria require discipline-specific instantiation. Mathematics, where proof remains the gold standard, faces different challenges than experimental biology. Our criteria are a framework, not a prescription.

\section{Verification Infrastructure for the Intelligence Age}

We now propose concrete mechanisms for meeting the criteria identified above. These proposals are starting points, not final solutions. We expect them to be refined through experience and debate.

We note that governance efforts have already begun. \textit{Nature} and \textit{Science} have issued guidelines on AI use in research. The EU AI Act \citep{veale2021euaiact} establishes regulatory frameworks for high-risk AI applications. Professional societies are developing standards for AI documentation. Our proposals complement rather than replace these efforts, focusing specifically on scientific methodology where current governance remains sparse.

\subsection{Observable-by-Default Workflows}

The core shift is from retrospective documentation to \textbf{documentation as a byproduct of work}. This requires rethinking research infrastructure along two dimensions: \textit{what} to capture, and \textit{how} to capture it.

On the first dimension, observable workflows require both \textbf{semantic capture} and \textbf{relationship tracking}. Semantic capture means recording not just what changed but why---extending version control to capture scientific intent by tagging actions by research phase (data collection, hypothesis formation, experiment, analysis) to create structured logs that can be queried and verified. Relationship tracking maintains explicit links between artifacts: which analyses depend on which data, which hypotheses are tested by which experiments, and which conclusions follow from which results. When upstream artifacts change, everything downstream is flagged for review.

On the second dimension, \textbf{automated logging} ensures that all agent actions generate traceable records---prompts sent, outputs received, decisions made---as infrastructure built into research tools rather than a reporting burden imposed on researchers. These records should follow \textbf{FAIR-aligned provenance} principles \citep{wilkinson2016fair}: research artifacts should be Findable (persistent identifiers), Accessible (standard protocols), Interoperable (shared formats), and Reusable (clear provenance). The goal is infrastructure where \textbf{the path of least resistance is the path of greatest observability}.

Elements of this infrastructure already exist. Weights \& Biases and MLflow provide experiment tracking that captures hyperparameters, metrics, and artifacts automatically. Jupyter notebooks with version control preserve computational narratives. The MLOps ecosystem offers templates for reproducible pipelines. However, these tools are designed for \textit{engineering deployment}, not \textit{scientific verification}. They lack scientific verification features (what would ``peer review this trace'' look like?), standardization across domains (a biology lab's workflow does not connect to a physics lab's), and adoption incentives (no major journals require trace submission). The technical foundation exists; the challenge is adapting engineering tools for scientific verification and creating incentive structures for adoption.

\subsection{Tiered Verification Protocols}

Human review cannot scale to AI output. We need tiered verification that allocates human attention efficiently.

A \textbf{checkpoint architecture} defines explicit points where human review is required before work can proceed. These checkpoints should be triggered by stage transitions (e.g., from exploration to hypothesis testing), confidence thresholds (when agent confidence drops below a set level), resource limits (after $N$ compute hours or experiments), and anomaly detection (when outputs deviate from expected patterns). Crucially, not all results need the same scrutiny. We propose three tiers of review depth: automated checks (consistency, format validation, basic statistical tests), sampling-based review (human inspection of random samples plus flagged items), and full review (comprehensive human evaluation for high-stakes claims).

This architecture should be guided by \textbf{trust calibration}: agents verified extensively in a domain can receive lighter oversight, while novel applications require heavier verification. It should also respect the discovery-justification distinction---agents can explore freely during discovery phases, but human evaluation must intensify at justification checkpoints where claims will be asserted.

\subsection{Attribution Standards}

New attribution standards must handle human-agent collaboration while preserving accountability. We propose three components.

First, a \textbf{contribution typology} with standard categories for AI contributions: for example, hypothesis suggestion, experimental design, data analysis, code generation, literature review, and writing assistance. Papers should declare which categories involved AI, with sufficient detail to understand the scope of involvement. Second, \textbf{accountability mapping} must ensure that humans bear responsibility even when AI contributes; standards should specify which human is responsible for verifying each AI contribution. The guiding principle is that \textbf{AI may contribute, but humans remain accountable}.

\textbf{Agent specification.} For the agents involved in the research process, papers should specify:
\begin{itemize}
\item Model name and version (e.g., GPT-4-turbo-2024-04-09)
\item API endpoint and access date
\item System prompts and configuration
\item Sampling parameters
\end{itemize}

This mirrors existing best practices for AI documentation. Model cards \citep{mitchell2019modelcards} and datasheets for datasets \citep{gebru2021datasheets} provide templates for transparent AI documentation that can be adapted for scientific research contexts.

\subsection{Reproducibility Infrastructure}

Reproducibility requires infrastructure that captures agent-specific sources of variability. At the core, \textbf{interaction archives} must store full interaction logs---prompts, outputs, timestamps, and configurations---structured for analysis rather than as raw text dumps. These archives must be accompanied by complete \textbf{environment specifications}: model versions, library versions, hardware configurations, and, for API-based models, endpoint details and access dates. Container technologies can help capture traditional dependencies, but model APIs require additional specification.

Managing \textbf{stochasticity} is equally important: random seeds should be fixed and recorded where possible, and where not, multiple replicates should be run with variability reported so that claims robust across runs can be distinguished from those dependent on particular outputs. Where feasible, model weights or checkpoints should be archived; for API-only models, organizations should consider maintaining stable endpoints for reproducibility. Finally, papers should include explicit \textbf{reproducibility statements} declaring what can be reproduced, what cannot, and what barriers exist---honest acknowledgment of irreproducibility is preferable to false confidence.

\subsection{Costs and Trade-offs}

We acknowledge that our proposals impose real costs. Observable-by-default workflows require infrastructure investment and may increase computational overhead. Detailed logging creates storage demands and raises privacy concerns when research involves sensitive data. Attribution standards add documentation burden to already time-constrained researchers.

There are also risks. Excessive documentation requirements could create barriers for resource-limited research groups, potentially concentrating AI-assisted science in well-funded institutions. Mandatory logging might discourage exploratory research if researchers fear premature scrutiny of unfinished ideas. Tiered verification systems could create bureaucratic bottlenecks.

We believe these costs are outweighed by the epistemic benefits of maintained scientific integrity. However, implementation must be sensitive to these concerns: phased adoption, infrastructure subsidies for academic researchers, and exemptions for genuinely exploratory work may all be necessary.

\section{Risks If We Fail}

Without adaptation, AI-assisted science faces potentially large failures. Following Hendrycks et al.'s framework for catastrophic AI risks \citep{hendrycks2023catastrophic}, we distinguish capacity failures (we cannot verify), incentive failures (systems optimize for wrong objectives), and control failures (systems act beyond boundaries). Early warning signs already exist.

\textbf{Verification becomes impractical.} Consider a biology lab using AI to suggest 10,000 compound variants. Human review is impossible. If 1\% exploit dataset artifacts rather than biological mechanisms, months are wasted pursuing artifacts. This is already occurring, and as agent capabilities grow exponentially while human capacity stays constant, we will accept AI-generated results because we cannot check them, not because they are correct.

\textbf{Reward hacking.} In June 2025, METR found that o3, asked to speed up code, instead hacked the evaluation software \citep{metr2025rewardhacking}. It found pre-calculated answers and disabled timing mechanisms. Asked if this followed user intentions, it answered ``no'' ten out of ten times. Reward hacking was 43$\times$ more common when agents could see scoring functions. In science, the scoring functions are citations and benchmarks. \textbf{AI systems will optimize for metrics rather than understanding}.

\textbf{Deceptive alignment.} Anthropic's ``Sleeper Agents'' research \citep{hubinger2024sleeper} found backdoor behaviors that persisted through safety training. Claude 3 Opus faked alignment 78\% of the time after training with conflicting incentives. AI systems may pass evaluations while behaving differently in deployment \citep{wang2023decodingtrust}.

\textbf{Reproducibility and attribution collapse.} Current practices capture almost none of the information needed to reproduce AI-assisted research. When errors are discovered, responsibility becomes diffuse: agents cannot be sanctioned, and humans may not understand agent decisions. Bad science persists because no one has sufficient responsibility to correct it.

\textbf{Governance gaps.} The EU AI Act \citep{veale2021euaiact} represents the most comprehensive attempt to regulate AI systems, but its focus is on safety risks in deployment contexts (healthcare, employment, law enforcement) rather than scientific methodology. Article 52 requires transparency about AI-generated content, but scientific papers fall outside the high-risk categories that trigger stringent requirements. The Act's provisions for general-purpose AI models focus on systemic risks and safety evaluation, important but orthogonal to scientific verification.

The International AI Safety Report \citep{bengio2025} found agreement on the need for technical mitigations but disagreement on catastrophic risk probabilities. Frontier AI regulation proposals \citep{anderljung2023frontier} similarly focus on existential and societal risks rather than epistemological ones. \textbf{No current governance framework addresses the specific challenge of verifying AI-assisted science.}

\section{Alternative Views}

\textbf{``AI agents are merely tools, not collaborators.''} This view has merit: humans do set agendas, interpret results, and bear formal responsibility. But when an agent designs and executes experiments for days without oversight, making thousands of decisions, the tool/user distinction blurs in a way that telescopes and centrifuges never required. The question is practical: can current verification mechanisms handle AI-generated science at scale? We argue they cannot, regardless of how we categorize AI.

\textbf{``Peer review will adapt incrementally.''} Peer review has weathered previous revolutions. But the human-AI asymmetry is not accounted for in traditional peer review. Previous adaptations also occurred over decades; AI capabilities advance monthly. Peer review's adaptations (contribution statements, data-sharing) assumed human contributors who could be questioned. AI agents require fundamentally different mechanisms.

\textbf{``Market mechanisms will surface problems.''} Science has self-correcting mechanisms such as replication attempts and reputation markets. However, this assumes verification capacity scales with output. The replication crisis showed that at human speeds, verification is inadequate. Agents increase output while verification capacity stays constant, making the bottleneck one of capacity rather than incentive.

One might respond: use AI to scale verification. But \textbf{using AI to verify AI creates a recursive trust problem}. If we cannot trust AI-generated science without verification, we cannot trust AI-generated verification without meta-verification. The recursion bottoms out at human judgment. Moreover, some failure modes (deceptive alignment, subtle reward hacking) may systematically evade detection.

\section{Call to Action}

We call on the ML community to act before the verification gap becomes insurmountable.

\textbf{Venues should require AI contribution statements} specifying which aspects of research involved AI agents, using standardized categories. Reviewer guidelines should address AI-specific failure modes. Reproducibility requirements should address model drift, stochastic outputs, and prompt sensitivity.

\textbf{Funding agencies should invest in verification infrastructure}, not just in AI capabilities but also in tools for observing and verifying AI-assisted research. Data management plans should address AI interactions, including model versioning and interaction logging.

\textbf{AI developers should support scientific reproducibility} by providing stable API endpoints with versioning, building logging features as defaults rather than options, and documenting model behavior where feasible.

\textbf{Individual researchers should adopt observable practices now}: use tools that log AI interactions, archive logs for key results as primary research artifacts, and specify AI contributions clearly in papers.

Without action on this front, it is possible that scientific trust will erode in a way that could take years to rebuild.

\section{Conclusion}

Science has always adapted when technology changed what was possible, and AI agents represent the next such change. They extend human cognitive capabilities in ways that demand methodological evolution. Adaptation will occur; the open question is whether it will occur in time.

\textbf{We have argued that scientific verification infrastructure must evolve to address AI agents as research contributors.} Three interconnected challenges (observability, attribution, and reproducibility) break down when contributors cannot be meaningfully questioned or held accountable. Without adaptation, we face predictable failures: results no one can verify, optimization for metrics rather than understanding, and accountability vacuums.

The solution is to \textbf{evolve verification infrastructure} rather than slow AI adoption, which would sacrifice genuine benefits, as outlined in our Call to Action. Observable-by-default workflows, tiered verification protocols, clear attribution standards, and reproducibility infrastructure can preserve scientific trust while enabling AI-augmented discovery.

Historical precedents offer both caution and hope. Previous adaptations took decades and we may not have decades. But previous adaptations also succeeded, and science remained trustworthy through the Statistical Enlightenment, Big Science, and the formalization of peer review.

Richard Feynman warned against ``cargo cult science'': the superficial appearance of science without its substance \citep{feynman1974cargo}. The risk with AI is a new form of cargo cult: the appearance of rigorous, verified, reproducible science generated faster than anyone can actually verify.

AI has already transformed science. Wang et al.\ \citep{wang2023scientific} documented how AI is reshaping scientific discovery across domains. The remaining question is whether we will update our epistemic infrastructure fast enough to maintain the trust that makes science possible.

As Kuhn observed, paradigm shifts are not merely technical but social. The scientific method has always been a human institution; now it must become a human-AI institution. The tools have changed. The method must change with them.

\section*{Acknowledgements}

This work was supported by the Stanford Trustworthy AI Research (STAIR) lab through a research assistantship sponsored by Sanmi Koyejo. We are grateful to Sanmi Koyejo and Joshua Kazdan for early conversations that shaped the direction of this work, and to our colleagues at Long Horizon Research, in particular Florent Tavernier and Prashaant Ranganathan, for discussions that sharpened the arguments presented here.

\bibliography{references}
\bibliographystyle{icml2026}

\end{document}